\documentclass[superscriptaddress, reprint, amsmath, amssymb, aps, prl]{revtex4-1}%
\usepackage{bm}%
\usepackage{graphicx}%
\usepackage{xspace}%
\usepackage{dcolumn}%
\usepackage{color}%

\newcommand{\CuO}{$\mathrm{CuO_2}$\xspace}%
\newcommand{\lco}{$\mathrm{La_2CuO_4}$\xspace}%
\newcommand{\lem}{LE-$\mu$SR\xspace}%
\newcommand{\muSR}{$\mu$SR\xspace}%
\newcommand{\ie}{\emph{i.e.\xspace}}%
\newcommand{\eg}{\emph{e.g.\xspace}}%
\newcommand{\HAF}{2DHAF\xspace}%
\newcommand{\SL}[1]{[3LSCO+#1LCO]\xspace}%
\newcommand{\etal}{\emph{et al.}\xspace}%
\begin{document}

%
\title{Magnetism in the 2D Limit and Interface Superconductivity in Metal-Insulator $\mathbf{La_{2-x}Sr_{x}CuO_{4}}$ Superlattices}%
\author{A. Suter}%
\email{andreas.suter@psi.ch}%
\affiliation{Laboratory for Muon Spin Spectroscopy, Paul Scherrer Institute, 5232 Villigen PSI, Switzerland}%
\author{E. Morenzoni}%
\affiliation{Laboratory for Muon Spin Spectroscopy, Paul Scherrer Institute, 5232 Villigen PSI, Switzerland}%
\author{T. Prokscha}%
\affiliation{Laboratory for Muon Spin Spectroscopy, Paul Scherrer Institute, 5232 Villigen PSI, Switzerland}%
\author{B. M. Wojek}%
\affiliation{Physik-Institut, Universit\"{a}t Z\"{u}rich, 8057 Z\"{u}rich, Switzerland}%
\affiliation{Laboratory for Muon Spin Spectroscopy, Paul Scherrer Institute, 5232 Villigen PSI, Switzerland}%
\author{H.~Luetkens}%
\affiliation{Laboratory for Muon Spin Spectroscopy, Paul Scherrer Institute, 5232 Villigen PSI, Switzerland}%
\author{G. Nieuwenhuys}%
\affiliation{Laboratory for Muon Spin Spectroscopy, Paul Scherrer Institute, 5232 Villigen PSI, Switzerland}%
\author{A. Gozar}%
\affiliation{Brookhaven National Laboratory, Upton, New York 11973-5000, USA}%
\author{G. Logvenov}%
\altaffiliation{present address: Max-Planck-Institut f\"{u}r Festk\"{o}rperforschung, 70569 Stuttgart, Germany}%
\affiliation{Brookhaven National Laboratory, Upton, New York 11973-5000, USA}%
\author{I. Bo\v{z}ovi\'{c}}%
\affiliation{Brookhaven National Laboratory, Upton, New York 11973-5000, USA}%
\date{\today}%
\begin{abstract}
We show, by means of low-energy muon spin rotation measurements, that few-unit-cells thick \lco layers synthesized digitally by molecular beam epitaxy synthesis are antiferromagnetically ordered. Below a thickness of about 5 \CuO layers the long-range ordered state breaks down, and a magnetic state appears with enhanced quantum fluctuations and a reduced spin stiffness. This magnetic state can exist in close proximity (few \AA) to high-temperature superconducting layers, without transmitting supercurrents.
\end{abstract}

\maketitle

By reducing the dimensionality of a solid, its electronic states and physical properties can be drastically modified, but these changes are not easy to predict for strongly correlated electron materials. For example, in thin interfacial layers inside oxide heterostructures a host of electronic states were discovered experimentally --– a high-mobility 2D electron gas \cite{ohtomo_high-mobility_2004}, magnetism \cite{brinkman_magnetic_2007}, quantum Hall effect \cite{tsukazaki_quantum_2007}, and interface superconductivity between insulators \cite{reyren_superconducting_2007}. In metal-insulator (MI) bilayer $\mathrm{La_{1.55}Sr_{0.45}CuO_4}$/\lco heterostructures (LSCO-LCO), where none of the constituents is superconducting, interface superconductivity with $T_{\rm c} \approx 30$~K has been discovered recently \cite{gozar_high-temperature_2008}. 

Up to now these studies used probes that are sensitive to charge but tell little about the microscopic magnetic state. For instance although it is known that 1 unit cell of LCO sandwiched between two optimally doped LSCO layers is still insulating \cite{bozovic_no_2003}, one can only speculate about its magnetic state. LCO is assumed to be close to the realization of a spin-$1/2$ isotropic Heisenberg antiferromagnet on a square lattice (\HAF), since its in-plane exchange constant, $J$, is about $10^4$ times larger than any other exchange coupling present. Bulk material shows antiferromagnetic (AF) long-range order (LRO) below a N\'{e}el temperature of $T_{\rm N} \simeq 310$ K \cite{yamada_spin_1989,maclaughlin_abrupt_1994}. In thin films reducing the thickness results in a decreased $T_{\rm N}$ \cite{suter_thinFilm_AF}, whereas strain seems to play only a minor role \cite{suter_antiferromagnetic_2004}. In the 2D limit, at any finite temperature LRO will be destroyed by thermal fluctuations \cite{mermin_absence_1966,hohenberg_existence_1967}. P.W. Anderson proposed that for the \HAF \cite{anderson_resonating_1987} even at $T = 0$ quantum fluctuations destroy LRO; instead, a quantum spin-liquid --– the resonating valence bond (RVB) state –-- should form. Chakravarty, Halperin, and Nelson \cite{chakravarty_two-dimensional_1989} solved the \HAF in the long wave limit and arrived at a different picture. The phase diagram is basically controlled by the temperature and the spin stiffness, $\rho_{\rm S}$, and only part of the phase diagram is dominated by quantum fluctuations (quantum disordered regime), whereas in the other part the spin correlation length, $\xi(T)$, grows exponentially by lowering the temperature (renormalized classical regime, RC). Indeed, measurements of $\xi(T)$ in the paramagnetic phase ($T>T_{\rm N}$) of bulk LCO revealed that it follows the RC behavior \cite{keimer_magnetic_1992,imai_low_1993}. While numerical simulations support the long-wave-limit calculations \cite{hirsch_two-dimensional_1985,loew_nel_2007}, it has been argued \cite{murthy_random_1988,read_valence-bond_1989,poilblanc_spinon_2006,sachdev_quantum_2008} that small deviations from the ideal \HAF, due to frustrating second-neighbor exchange, charge carrier doping, defects, etc., could reduce $\rho_{\rm S}$ and thus enhance the effect of quantum fluctuations, preventing the spins from acquiring LRO.

In this Letter we present a study focusing on the magnetic state of LCO layers within MI LSCO-LCO superlattices (SLs), where the number of \CuO layers within the LCO stack can be varied to approach the 2D limit. To probe AF order and magnetic fluctuations we used polarized low-energy muons as a local probe. Low-energy muon spin rotation (\lem) \cite{morenzoni_nano-scale_2004} can detect superconductivity and/or magnetism, either static or fluctuating, even in ultrathin layers \cite{luetkens_observation_2003}. We show that down to about 5 \CuO layers LCO acquires LRO at low enough temperatures. Below this thickness, LCO enters a different magnetic state, characterized by short-range correlations, and increased magnetic fluctuations. This indicates a cross-over to a quantum disordered regime in this \HAF model system. Furthermore, we show that this magnetic state exists in close spatial proximity to superconducting layers.

We have synthesized and studied a series of samples that contain ultrathin, isolated layers of LCO. The synthesis was carried out by means of an atomic-layer-by-layer molecular beam epitaxy (ALL-MBE) system equipped with in-situ surface science tools. ALL-MBE allows for synthesis of complex heterostructures in which the thickness of individual layers can be controlled down to a single atomic layer \cite{bozovic_no_2003,logvenov_high-temperature_2009}. We digitally varied the thickness of LCO layers alternating with metallic $\mathrm{La_{1.56}Sr_{0.44}CuO_4}$ (LSCO) layers. Counting in $1/2$-unit-cell (UC) increments, each of which contains a single \CuO plane, the investigated SLs have the repeat structure \SL{6}, \SL{9}, and \SL{12}, respectively. All SLs were grown on $\mathrm{LaSrAlO_4}$ substrates. The total film thickness was kept at about 85 nm. The lattice parameters of the SLs obtained by X-ray diffraction are:
\SL{6} $a/c=3.796$ \AA $/13.232$ \AA, \SL{9} $3.798$ \AA $/13.223$ \AA, \SL{12} $3.799$ \AA $/13.220$ \AA. If there is a difference in the LCO/LSCO sub-units of the SL, they couldn't be resolved. The trend of the lattice parameters is consistent with single phase thin films \cite{locquet_variation_1996}. A detailed analysis revealed that Sr interdiffusion is limited to about 1 unit cell thickness \cite{gozar_high-temperature_2008}. Resonant soft X-ray scattering was used \cite{smadici_superconducting_2009} to measure the charge redistribution along the $c$-axis in LSCO-LCO superlattices, indicating a characteristic screening length of about 0.6\, nm. Zn delta-doping tomography on bi-layers confirmed this and further revealed that the first \CuO plane from the interface on the LCO side is overdoped, the second one nearly optimally doped, and the third one heavily underdoped \cite{logvenov_high-temperature_2009}. Such charge redistribution is in fact expected from simple model calculations \cite{logvenov_high-temperature_2009,loktev_model_2008}. Mutual induction measurements confirmed that all investigated SLs have a superconducting transition temperature $T_c \simeq 25$\,K.

\begin{figure}[h]
  \centering
  \includegraphics[width=\linewidth]{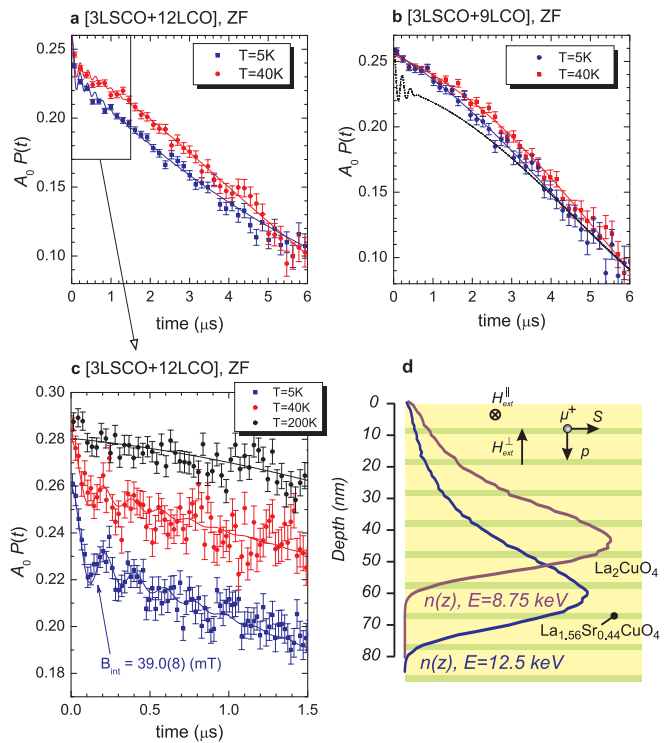}
  \caption{(a)--(c) $\mu^+$ spin-polarization spectra, $A_0 P(t)$, for zero 
           applied magnetic field. (a) $A_0 P(t)$ for the \SL{12} SL. 
           The $T=40$ K spectrum (upper curve, red online) is Gaussian like, whereas the 
           $T=5$ K (lower curve, blue online) is more exponential-like, which indicates 
           enhanced spin dynamics at lower temperatures.
           (b) $A_0 P(t)$ for the \SL{9} SL. Here $A_0 P(t)$ shows only a very weak additional
           exponential contribution at $T=5$ K compared to \SL{12}, and no zero-field precession is
           observable. For details see the text. (c) enlarged scale and different binning for \SL{12} 
           showing zero-field precession signals, \ie a well defined static internal field at the muon 
           site. The $T=40$ K and $T=200$ K curves are shifted up by 0.02 for clarity. 
           (d) The $\mu^+$ stopping distributions $n(z)$, used in the experiments. The yellow
           stripes represents the LCO, the green ones the LSCO within the SL, shown for \SL{12}.}\label{fig:LSCO-SL-ZF} 
\end{figure}

To detect magnetism we performed \muSR experiments as a function of temperature under zero-field conditions (ZF). To quantify the magnetic volume fraction and the robustness of the magnetic state, as well as to characterize superconductivity we applied small magnetic fields parallel and perpendicular to the $ab$-planes, always perpendicular to the muon spin (``transverse field'', TF). For each muon spin rotation measurement, a mosaic of four nominally identical $1\times1$ cm$^2$ samples was used. Fig.\ref{fig:LSCO-SL-ZF}d shows the muon stopping distributions as used in the experiments. The time evolution of the polarization of the muon ensemble $A_0 P(t)$, which is obtained form the muon decay spectra (typically from a few million muon decays) yields information about local magnetic field distributions at the muon stopping site and their static and dynamic properties. In case of AF LRO, muon spins precess in the internal field, $B_{\rm int}$, of the electronic magnetic moments with a frequency, $\nu_\mu=(\gamma_\mu/2\pi)\,B_{\rm int}$ ($\gamma_\mu$ is the gyromagnetic ratio of the muon), proportional to the staggered magnetization, and oscillations at this frequency show up in the polarization spectra. The presence of substantial magnetic disorder (\textit{e.g.}, a frozen spin glass state) or electronic low-frequency ($<10$ MHz) fluctuations leads to a strongly damped $A_0 P(t)$ due to a rapid dephasing of the muon spin ensemble. However, electronic high-frequency fluctuations ($>100$ MHz) will only lead to a weak depolarization of $A_0 P(t)$ (motional narrowing regime \cite{abragam_principles_1983}).

ZF polarization spectra are shown in Figs. \ref{fig:LSCO-SL-ZF}a and \ref{fig:LSCO-SL-ZF}b for \SL{12} and \SL{9}, respectively. In \SL{12} we observe static AF order, evident from the spontaneous zero-field precession signals (Fig. \ref{fig:LSCO-SL-ZF}c). The internal field, $B_{\rm int} = 39.0(8)$ mT, is equal to what is observed in single-phase LCO films \cite{suter_thinFilm_AF} and bulk samples \cite{borsa_staggered_1995}, thus showing that the full electron magnetic moment is present. The magnetic volume fraction estimated from the oscillatory amplitude of $A_0 P(t)$ shows that about $1/4$ of the film volume is magnetically ordered. From this, we can estimate the magnetic layer thickness to be $d_{\rm mag} \approx 3$ nm (4--5 \CuO layers), which is in quantitative agreement with simple model calculations taking into account Sr interdiffusion and charge redistribution between the M and I layers \cite{loktev_model_2008,gozar_high-temperature_2008}. According to the model, the inner 5 \CuO layers have doping levels of $x < 0.006$ and lie well within the AF part of the phase diagram ($T_{\rm N}\to 0$ for $x \gtrsim 0.02$). The model also predicts that one or two nominally insulating \CuO planes at the interface will have doping levels corresponding to the superconducting part of the LSCO phase diagram.

Our measurements of the superconducting properties of these SLs by \lem provide a further independent confirmation of the presence of interface superconductivity and of the charge levels in the SLs. From the absence of Meissner screening of a magnetic field applied parallel to the SL ($ab$-planes) we infer that no supercurrents flow along the $c$-axis, consistent with superconductivity being restricted to the interface. The London penetration depth, $\lambda_{\rm L}$, was estimated from the increased muon depolarization rate below $T_{\rm c}$ when field-cooling the sample in a field applied perpendicular to the $ab$-planes. Assuming a pancake vortex model \cite{clem_two-dimensional_1991,brandt_properties_2003} which takes into account the layered structure of the SLs, we find $\lambda_{\rm L}\approx 350\,\mathrm{nm}$ which is about 1.5 times larger than $\lambda_{\rm L}$ of optimally doped bulk LSCO \cite{Luke_magnetic_97}. Using the clean limit relation for the superfluid density 
$n_{\rm S} = \frac{1}{\mu_0}\, \frac{m}{e^2} \frac{1}{\lambda_{\rm L}^2}$, we find an averaged superfluid density of about half the value of optimally doped LSCO, again in satisfactory agreement with the charge transfer model. 

In contrast to the \SL{12} SL no signs of spontaneous ZF precession --- and hence no evidence for static AF LRO --- are found down to $T=5$\,K in \SL{9} and \SL{6}. Fig.\,\ref{fig:LSCO-SL-ZF}b shows the ZF time spectra for the \SL{9} sample. At $T=40$\,K, $A_0 P(t)$ shows a Gaussian depolarization typical for nuclear dipole fields. At $T=5$\,K an additional very weak exponential component appears. The dash-dotted line shows the expected $A_0 P(t)$, assuming a doping level $x$ calculated from the charge redistribution model and using experimental parameters from measurements on single-phase films with the corresponding $x$ values. Clearly, the predicted and measured spectra differ in two major features: the experimental data show no spontaneous precession and no sign of a fast initial depolarization is visible. The former points to the absence of static LRO, and the latter indicates that even static disordered magnetism, which would lead to a fast depolarization, is significantly suppressed.

\begin{figure}
 \includegraphics[width=0.9\linewidth]{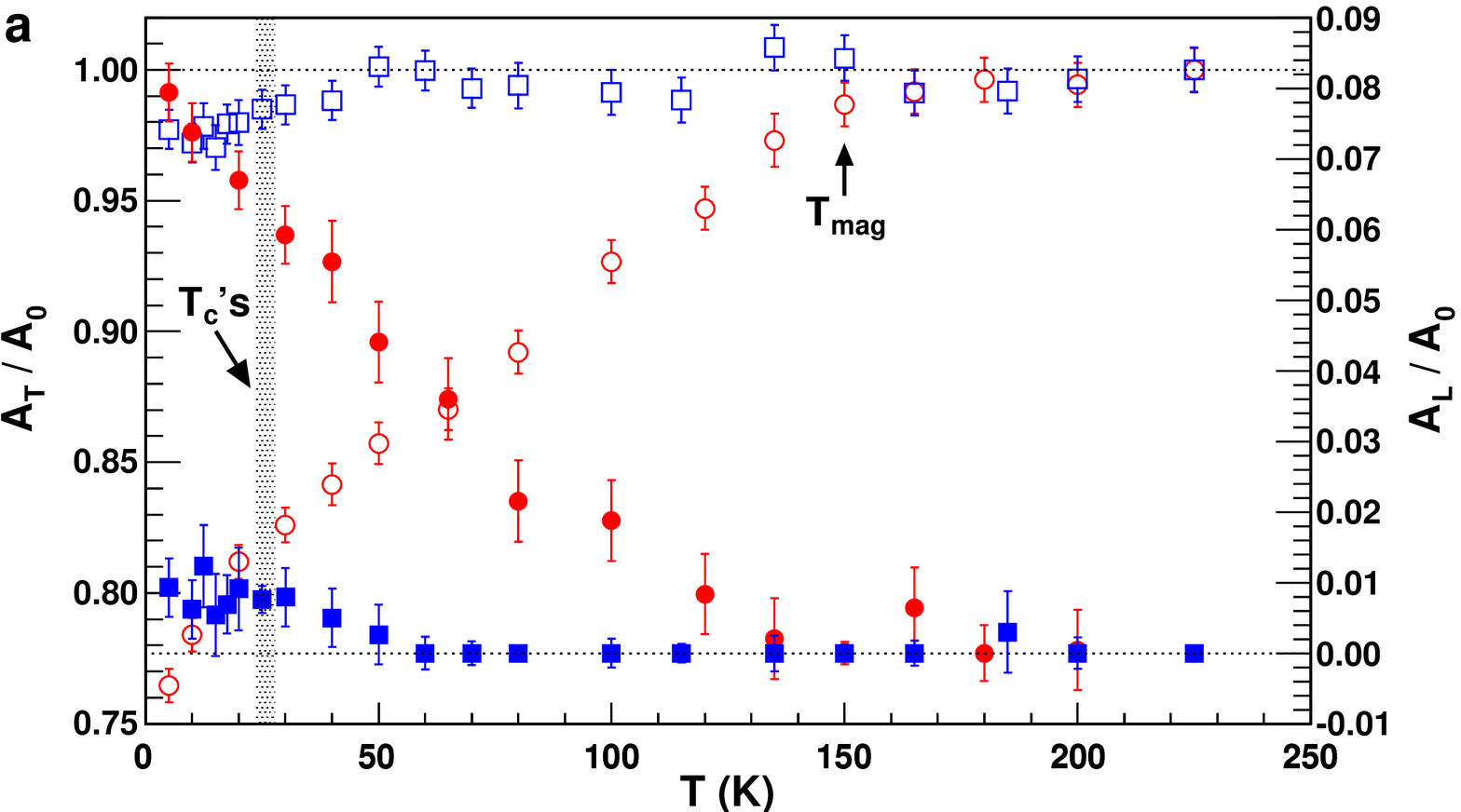}\\
 \includegraphics[width=0.9\linewidth]{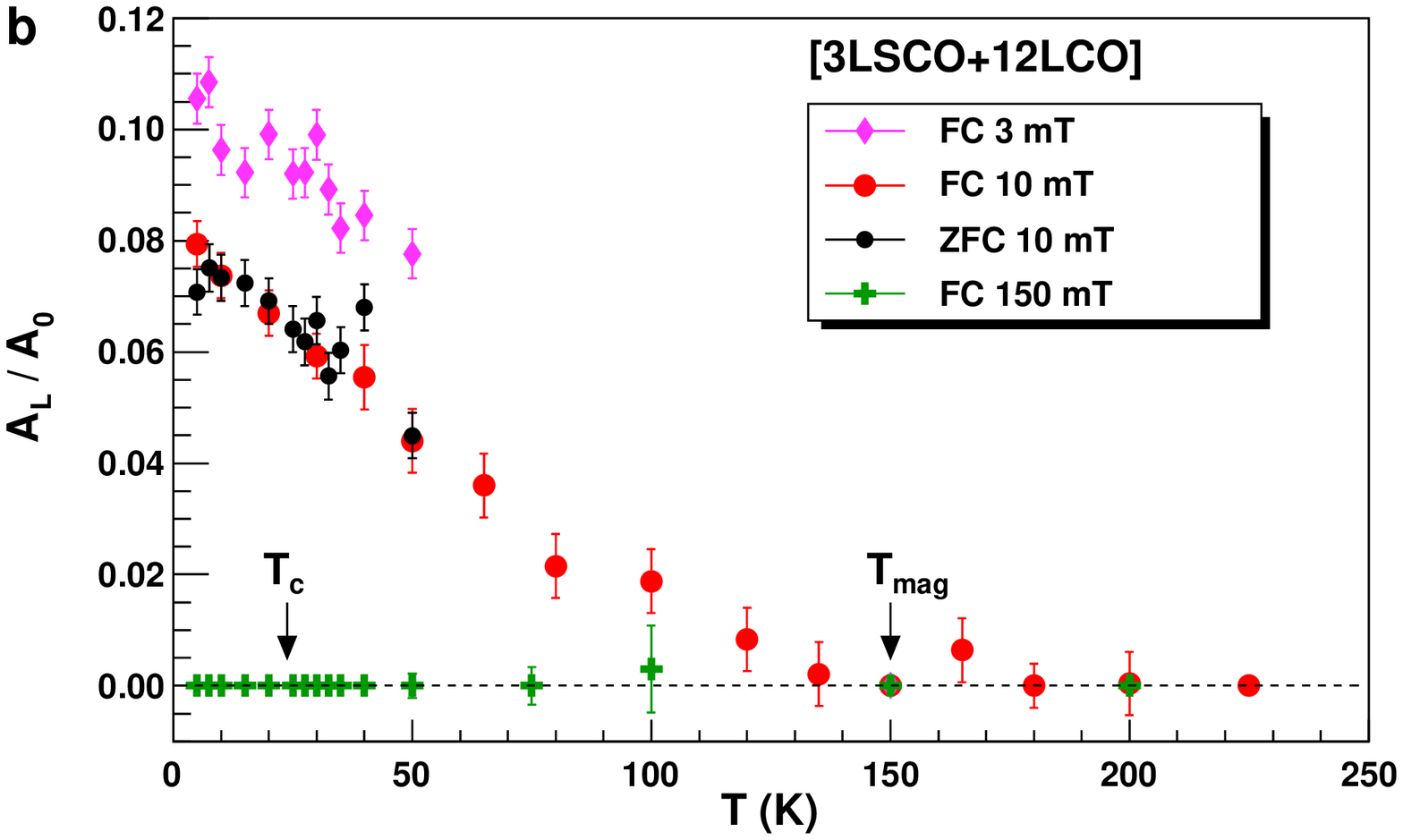}\\ 
 \caption{(a) Normalized transverse ($A_{\rm T}/A_0$) and longitudinal ($A_{\rm L}/A_0$) asymmetry as function of 
          temperature for the different SLs, obtained from weak transverse field measurements in 
          $B_{\rm ext}=10\,\mathrm{mT}$. Open symbols belong to left, closed symbols to the right axis. Red symbols: \SL{12}, 
          blue symbols: \SL{9}. Note: for non-magnetic samples $A_{\rm L} \equiv 0$. The shaded area shows the 
          region where the superconductive transition takes place: \SL{9}, $T_{\rm c} = 25.0\,\mathrm{K}$,
          \SL{12}, $T_{\rm c} = 24.0\,\mathrm{K}$. (b) $A_{\rm L}/A_0$ versus temperature for different 
          external fields $B_{\rm ext}$ for the \SL{12} SL, for field cooling (FC) and zero-field 
          cooling (ZFC). The pronounced reduction of $A_{\rm L}/A_0$ between 3 mT and 10 mT is
          due to a strong reduction of the spin stiffness compared to bulk LCO.}\label{fig:LSCO-SL-AT-AL}
\end{figure}

To further investigate the magnetic state we estimated the magnetic volume fraction in the samples by measuring  $A_0 P(t)$ in a weak magnetic field applied transverse to the initial muon polarization. In this case $A_0 P(t)$ can be written as \cite{pratt_field_2007}:

\begin{equation}\label{eq:wTF}
 A_0 P(t) = A_{\rm T} \exp[-(\sigma t)^2/2] \cos(\gamma_\mu B_{\rm tot} t + \phi) + 
            A_{\rm L} e^{-\lambda t} \cos(\phi),
\end{equation}

\noindent where $A_{\rm T}$, $\sigma$ and $A_{\rm L}$, $\lambda$ are the asymmetries and corresponding depolarization rates, transverse and parallel to the total field $B_{\rm tot} = | \langle \mathbf{B}_{\rm ext} + \mathbf{B}_{\rm int} \rangle |$, while $\phi$ is the detector phase. $\lambda$ was negligibly small in all measurements. $A_{\rm L}$ is a measure of the presence of static magnetism (ordered or disordered) and its volume fraction. For instance, in the case of static magnetic order with an underlying isotropic magnetic field distribution and 100\% volume fraction $A_{\rm L}/A_0$ will grow to $1/3$. In contrast, $A_{\rm T}/A_0$ would drop to zero in the magnetic phase. In any para- or diamagnetic sample, $A_{\rm L}$ will be identically zero at all temperatures. The resulting magnetic layer thicknesses can be estimated from the relation $d_{\rm mag} \approx (3+n) (c/2) (1-A_{\rm T}/A_0)$, where $n=6,\,9,\,12$ depending on the SLs, and $A_{\rm T}/A_0$ is shown in Fig.\,\ref{fig:LSCO-SL-AT-AL}a. The estimated magnetic layer thicknesses for $n=6,9$ is $d_{\rm mag} \approx$ 0.4--1 nm ($\approx$ 1--2 \CuO layers), again in agreement with the calculated charge distribution (inner layer doping estimate: \SL{9}  $x<0.008$, \SL{6} $x<0.025$). Since the inner layer doping of the \SL{6} is at the border of the AF region, it will not be discussed here.

For \SL{9} $A_{\rm L}$ is drastically reduced, and both $A_{\rm L}/A_0$ and $1-A_{\rm T}/A_0$ show only a small deviation from zero below 50 K. This behavior is typical for fast fluctuations where $A_{\rm L}$ is vanishing at all temperatures. We ascribe this difference, together with the absence of a ZF precession and the very weak initial drop of $A_0 P(t)$ (Fig.\,\ref{fig:LSCO-SL-ZF}b), to increased fluctuations in ultrathin LCO layers, which prevent the formation of either LRO or a static disordered magnetic state (\eg spin glass).

These fluctuations are not expected within the RC regime. The following estimate indicates that they are of quantum nature. Within the RC regime $\xi$ is given as: $\xi(T)/a = 0.5 \exp(1/y) \left[ 1 - y/2 + \mathcal{O}(y^2) \right]$, with $a$ the in-plane lattice constant, $y = k_{\rm B}T/(1.13 J)$, and $J/k_{\rm B}\approx 1500$ K for LCO. At $T\approx 150$ K, $\xi/a > 10^4$ which should result in a quasi-static magnetic state, \ie either ZF precession or, in the strongly disordered case, a strong initial depolarization should be observable. Both are absent in the \SL{9} and \SL{6} SLs down to the lowest temperature. The same conclusion is reinforced by the fact that the time-independent component
$A_{\rm L}/A_0$ is drastically reduced, when decreasing the number of \CuO planes in LCO (Fig. \ref{fig:LSCO-SL-AT-AL}b) and by application of increasing magnetic fields (Fig. \ref{fig:LSCO-SL-AT-AL}c). Assuming a random static internal field within the \CuO planes, the magnetic field dependence of $A_{\rm L}(b) \propto 1/(2 [1+b^2])$ with $b = B_{\rm ext}/B_{\rm int}$. The expected ratio $R\equiv A_{\rm L}(B_{\rm ext, 2})/A_{\rm L}(B_{\rm ext, 1})|_{T\to 0}$ for $B_{\rm ext, 1} = 3\,\mathrm{mT}$, and $B_{\rm ext, 2} = 10\,\mathrm{mT}$ and the measured internal field of $B_{\rm int} = 39\,\mathrm{mT}$ is $R=0.94$, however, for the \SL{12} a value of $R=0.76(1)$ is found (see Fig. \ref{fig:LSCO-SL-AT-AL}c). This drastic reduction can only originate from fluctuations and cannot be due to disorder. In order to see if unexpected doping, \ie deviations from the simple charge-transfer model, could lead to such a strong modification of the magnetic state, we performed the same measurements on 53\,nm thick single phase $\mathrm{La_{1.97}Sr_{0.03}CuO_4}$ films and find $R=0.95(2)$ in the so-called cluster-spin glass phase. This is in excellent agreement with the static model estimate, indicating that the strong reduction of $R$ in the SLs is due to dimensional effects, \ie increased magnetic fluctuations, and not due to charge-transfer effects or disorder. 

Another estimate further supports our finding: from the known magnon dispersion \cite{coldea_spin_2001} in LCO one can put a lower limit on the magnon wavelength that can be thermally excited at $T = 5$ K to about $1~\mu\mathrm{m}$, and this is the length scale on which magnons would destroy static long-range AF order. Experimentally, we see the absence of LRO on the length scale of less than 5\, nm (local order on this length scale would lead to ZF muon spin precession or strong damping), this requires the presence of very short-wavelength, high-energy ($\sim 100$ meV) AF fluctuations, which at $T = 5$ K can only be of quantum nature.

All these findings show that LCO within these SLs is not in the RC regime (as is the case for bulk LCO for $T>T_{\rm N}$), \ie the spin stiffness of the AF state is drastically reduced. Currently we do not know what is the reason for this strong reduction of the spin stiffness, however we can rule out that it is caused by disorder.


Acknowledgments: The \muSR experiments were fully performed at the S$\mu$S.
The work at BNL was supported by the U.S. Department of Energy Project MA-509-MACA.


\begin{thebibliography}{39}%
\bibitem{ohtomo_high-mobility_2004}%
  A. Ohtomo and H. Y. Hwang, Nature \textbf{427}, 423 (2004).

\bibitem{brinkman_magnetic_2007}%
  A. Brinkman, \etal, Nature Materials \textbf{6}, 493 (2007).

\bibitem{tsukazaki_quantum_2007}%
  A. Tsukazaki, \etal, Science \textbf{315}, 1388 (2007).

\bibitem{reyren_superconducting_2007}%
  N. Reyren, \etal, Science \textbf{317}, 1196 (2007).

\bibitem{gozar_high-temperature_2008}%
  A. Gozar, \etal, Nature \textbf{455}, 782 (2008).

\bibitem{bozovic_no_2003}%
  I. Bo\v{z}ovi\'{c}, \etal, Nature \textbf{422}, 873 (2003).

\bibitem{yamada_spin_1989}%
  K. Yamada, \etal, Phys. Rev. B \textbf{40}, 4557 (1989).

\bibitem{maclaughlin_abrupt_1994}%
  D. MacLaughlin, \etal, Phys. Rev. Lett. \textbf{72}, 760 (1994).

\bibitem{suter_thinFilm_AF}%
  A. Suter, \etal, unpublished.

\bibitem{suter_antiferromagnetic_2004}%
  A. Suter, \etal, J. Magn. Magn. Mater. \textbf{272-276}, 110 (2004). 

\bibitem{mermin_absence_1966}%
  N. Mermin and H. Wagner, Phys. Rev. Lett. \textbf{17}, 1133 (1966). 

\bibitem{hohenberg_existence_1967}%
  P. Hohenberg, Phys. Rev. \textbf{158}, 383 (1967).

\bibitem{anderson_resonating_1987}%
  P. W. Anderson, Science \textbf{235}, 1196 (1987).

\bibitem{chakravarty_two-dimensional_1989}%
  S. Chakravarty, B. Halperin, and D. Nelson, Phys. Rev. B \textbf{39}, 2344 (1989).

\bibitem{keimer_magnetic_1992}%
  B. Keimer, \etal, Phys. Rev. B \textbf{46}, 14034 (1992). 

\bibitem{imai_low_1993}%
  T. Imai, C. Slichter, K. Yoshimura, and K. Kosuge, Phys. Rev. Lett. \textbf{70}, 1002 (1993).
  T. Imai, C. Slichter, K. Yoshimura, M. Katoh, and K. Kosuge, \emph{ibid.} \textbf{71}, 1254 (1993).

\bibitem{hirsch_two-dimensional_1985}%
  J. Hirsch, Phys. Rev. B \textbf{31}, 4403 (1985).

\bibitem{loew_nel_2007}%
  U. L\"{o}w, Phys. Rev. B \textbf{76}, 220409 (2007).

\bibitem{murthy_random_1988}%
  G. Murthy, Phys. Rev. B \textbf{38}, 5162 (1988).

\bibitem{read_valence-bond_1989}%
  N. Read and S. Sachdev, Phys. Rev. Lett. \textbf{62}, 1694 (1989).

\bibitem{poilblanc_spinon_2006}%
  D. Poilblanc, A. L\"{a}uchli, M. Mambrini, and F. Mila, Phys. Rev. B \textbf{73}, 100403 (2006).

\bibitem{sachdev_quantum_2008}%
  S. Sachdev, Nature Physics \textbf{4}, 173 (2008).

\bibitem{morenzoni_nano-scale_2004}%
  E. Morenzoni, \etal, J. Phys.: Condens. Matter \textbf{16}, S4583 (2004).
  T. Prokscha, \etal, Nucl. Instrum. Methods Phys. Res., Sect. A \textbf{595}, 317 (2008).

\bibitem{luetkens_observation_2003}%
  H. Luetkens, \etal, Phys. Rev. Lett. \textbf{91}, 017204 (2003).
  A. Suter, \etal, \emph{ibid.}, \textbf{92}, 087001 (2004).
  E. Morenzoni, \etal, \emph{ibid.}, \textbf{100}, 147205 (2008).

\bibitem{logvenov_high-temperature_2009}%
  G. Logvenov, A. Gozar, and I. Bo\v{z}ovi\'{c}, Science \textbf{326}, 699 (2009).

\bibitem{locquet_variation_1996}%
  J.-P. Locquet, \etal, Phys. Rev. B \textbf{54}, 7481 (1996). 

\bibitem{smadici_superconducting_2009}%
  S. Smadici, \etal, Phys. Rev. Lett. \textbf{102}, 107004 (2009).

\bibitem{loktev_model_2008}%
  V. Loktev and Y. Pogorelov, Phys. Rev. B \textbf{78}, 180501 (2008). 

\bibitem{abragam_principles_1983}%
  A. Abragam, \textit{The principles of nuclear magnetism} (Oxford University Press, 1983).

\bibitem{borsa_staggered_1995}%
  F. Borsa, \etal, Phys. Rev. B \textbf{52}, 7334 (1995).

\bibitem{clem_two-dimensional_1991}%
  J. Clem, Phys. Rev. B \textbf{43}, 7837 (1991).

\bibitem{brandt_properties_2003}%
  E. H. Brandt, Phys. Rev. B \textbf{68}, 054506 (2003).

\bibitem{Luke_magnetic_97}%
  G. M. Luke, \etal, Physica C \textbf{282-287}, 1465 (1997). 

\bibitem{pratt_field_2007}%
  F. Pratt, J. Phys.: Condens. Matter \textbf{19}, 456207 (2007).

\bibitem{coldea_spin_2001}%
  R. Coldea, \etal, Phys. Rev. Lett. \textbf{86}, 5377 (2001).


\end{thebibliography}

%

\end{document}